\documentclass[aps,twocolumn,preprintnumbers,longbibliography]{revtex4-1}

\usepackage{graphicx}  % Include figure files
\usepackage{subfigure}
\usepackage{multirow}

\usepackage{fancyhdr}
\usepackage{longtable}
\usepackage{parskip}
\usepackage[T1]{fontenc}
\usepackage{dcolumn}   % Align table columns on decimal point

\usepackage{bm}        % bold math
\usepackage{amsfonts}  % Common math fonts
\usepackage{amsmath}   % Common math functions
\usepackage{amssymb}   % Common math symbols
\usepackage{appendix}

\usepackage[colorlinks=true, filecolor=blue, citecolor=blue,urlcolor=blue]{hyperref}

\usepackage{tikz}      % math illustration lib
\usetikzlibrary{arrows.meta}
\usetikzlibrary{decorations.markings}
\usetikzlibrary{calc}
\usepackage{xcolor}  %color lib
\usepackage{colortbl}

\newcommand{\eq}{\begin{eqnarray}}
\newcommand{\en}{\end{eqnarray}}

\newcommand{\E}{{\bf E}}
\newcommand{\B}{{\bf B}}
\newcommand{\A}{{\bf A}}
\newcommand{\J}{{\bf J}}

\newcommand{\nvec}{{\text{\bf{n}}}}
\newcommand{\inv}{{\text{inv}}}
\newcommand{\vis}{{\text{vis}}}
\newcommand{\emt}{{\text{em}}}
\newcommand{\rec}{{\text{rec}}}
\newcommand{\eff}{{\text{eff}}}

\newcommand{\sign}{{\text{sign}}}

\newcommand{\bfnabla}{\bm{\nabla}}
\newcommand{\kA}{k_{A'}}

\newcommand{\mA}{m_{A'}}

\newcommand{\lvec}{{\text{\bf{l}}}}
\newcommand{\xvec}{{\text{\bf{x}}}}
\newcommand{\yvec}{{\text{\bf{y}}}}
\newcommand{\nn}{\nonumber}
\usepackage{comment}

\begin{document}
\preprint{INR-TH-2024-002}
\title{Light-shinning-through-thin-wall radio frequency cavities for probing dark photon}

\author{Dmitry~Salnikov}
\email[\textbf{e-mail}: ]{salnikov.dv16@physics.msu.ru}
\thanks{corresponding author}
\affiliation{Moscow State University, 119991 Moscow, Russia}
\affiliation{Institute for Nuclear Research, 117312 Moscow, Russia}

\author{D.~V.~Kirpichnikov}
\email[\textbf{e-mail}: ]{dmbrick@gmail.com}
\thanks{corresponding author}
\affiliation{Institute for Nuclear Research, 117312 Moscow, Russia}

\author{Petr~Satunin}
\email[\textbf{e-mail}: ]{petr.satunin@gmail.com}
\affiliation{Moscow State University, 119991 Moscow, Russia}
\affiliation{Institute for Nuclear Research, 117312 Moscow, Russia}

\author{Leysan~Valeeva}
\email[\textbf{e-mail}: ]{valeeva.ln19@physics.msu.ru}
\affiliation{Moscow State University, 119991 Moscow, Russia}
\affiliation{Institute for Nuclear Research, 117312 Moscow, Russia}

\date{\today}

\begin{abstract}  
 We address the radio frequency (RF) cavity experiment for probing dark photons, which is a modification of the light-shining-through-thin-wall (LSthinW) setup with a relatively thin conducting barrier between a cylindrical emitter and a hollow receiver. 
 The experimental facility allows for the effective probing of dark photons even in the off-shell regime, i.e., when the dark photon mass exceeds the driving frequency of the emitter cavity, which is pumped by an electromagnetic mode.  
 We compare the sensitivity of two specific setup configurations: (i) two adjacent cylindrical cavities placed end-to-end with an end-cap separating them, and (ii) a nested geometry in which the cylindrical receiver is encapsulated within the emitter. 
 We demonstrate that, for a certain range of dark photon masses, the nested configuration with the $\mbox{TM}_{010}$ pump mode can provide enhanced sensitivity compared to an adjacent emitter setup.
 Remarkably, for the $\mbox{TE}_{011}$ pump mode, both the nested and adjacent cavity configurations can yield comparable expected reaches for the specific geometry type.
 \end{abstract}

\maketitle 

\section{Introduction}

A dark (hidden) photon  is a massive hypothetical particle that interacts with the Standard Model (SM) through kinetic mixing with a visible photon~\cite{Holdom:1985ag}. 
For relatively small values of the kinetic mixing parameter, dark photon (DP) can avoid cosmological constraints, and thus it can be a viable candidate for dark matter~\cite{Nelson:2011sf,Arias:2012az}.
It is worth noticing that a hidden photon is addressed also in the scenarios as a mediator between dark matter and the visible sector~\cite{Antel:2023hkf,Zhevlakov:2023wel,NA64:2023wbi}. 
Typically, a DP is  described in the  framework of $U(1)$ gauge extension of the SM,  that is   a natural  requirement  for a  string  theory~\cite{Okun:1982xi}. 

A variety of precise searches for hidden photons have been carried out recently in numerous laboratory and accelerator based experiments. 
For instance, a search for hidden photons with relatively heavy masses  (in the $\lesssim \mathcal{O}(1)~\mbox{GeV}$ range) has utilized $e^+e^-$~\cite{BaBar:2017tiz} and $pp$ colliders~\cite{LHCb:2017trq}, the  NA64 fixed target~\cite{NA64:2019auh,NA64:2018lsq}, the JUNO underground detectors~\cite{DEramo:2023buu}, and  FASER beam-dump~\cite{FASER:2023tle} experiments. 

In addition, we also address the low mass limits (sub~$\sim \mathcal{O}(1)~\mbox{eV}$ range) derived from tests of Coulomb’s law~\cite{Williams:1971ms,Abel:2008ai} and atomic spectroscopy~\cite{Jaeckel:2010xx}. 
The limits on dark photons that can be hypothetical DM candidates are performed in haloscope searches ~\cite{Gelmini:2020kcu,BREAD:2021tpx,Chiles:2021gxk,Chen:2024aqf,Knirck:2023jpu,Chaudhuri:2014dla,Alesini:2023qed,McAllister:2022ibe,Tang:2023oid}, gravitational wave detectors~\cite{Ismail:2022ukp}, cryogenic millimeter-wave cavity~\cite{DOSUE-RR:2022ise}, dish antenna~\cite{Horns:2012jf}, radio telescopes~\cite{An:2020jmf}, gaseous detectors~\cite{Graesser:2024cns}, photonic chip~\cite{Blinov:2024jiz}, astrophysical environments~\cite{Brahma:2023zcw}, and Earth-based magnetometers~\cite{Bloch:2020lyz,Fedderke:2021aqo} (for recent review see also Ref.~\cite{Berlin:2022hfx} and references therein). 

For  completeness, it is worth noting that dark photons for the sufficiently small masses below  $\lesssim 1~\mbox{eV}$  have been probed with the SM photon regeneration facilities such as DarkSRF~\cite{Romanenko:2023irv}, CROWS~\cite{Betz:2013dza}, ADMX~\cite{ADMX:2010ubl}, and microwave cavity experiment~\cite{Povey:2010hs}. 

The latter facilities are referred to as light-shining-through-wall (LSW) experiments. 
To be more specific, the regarding  detection schemes to search for dark photons are associated with electromagnetically isolated conducting cavities tuned to the same frequency~\cite{VanBibber:1987rq,Jaeckel:2007ch}. 
In particular,  dark photons produced from high-density SM photons in the emitter cavity  propagate in space until reaching the other receiver  hollow  cavity, where they convert back into SM photons. 

\begin{figure}[tbh!]\centering
\includegraphics[width=0.857\linewidth]{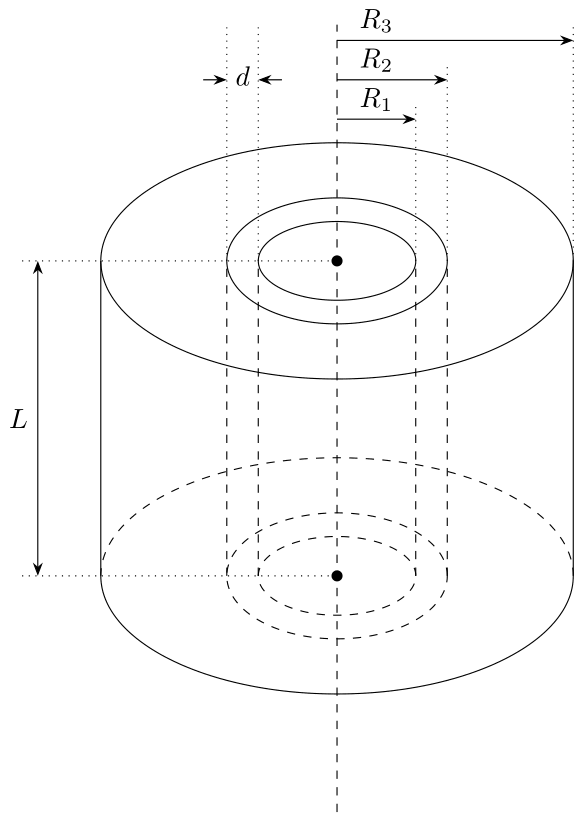}
\caption{ A typical sketch of the nested setup. 
The emitter cavity is considered a cylindrical layer with outer and inner radii $R_2$ and $R_3$, respectively.
The receiver is an internal hollow cylindrical cavity with radius $R_1$ that is encapsulated within the emitter. Both the outer and inner cavities have the same length $L$.
%cavity.
}
\label{fig:EncDesign}
\end{figure}

The authors of Ref.~\cite{Graham:2014sha} argued that, while previous literature~\cite{Betz:2013dza,Povey:2010hs,Jaeckel:2007ch,Kim:2020ask,Parker:2013fxa,Parker:2013fba} focused on the  production and detection of transverse modes of hidden photon, the longitudinal mode allows a significant improvement in LSW experimental sensitivity.
%В трёх предложениях ниже три ссылки на Берлина. Возможно, стоит одну оставить, или норм?
In the present paper, we employ the formalism of Ref.~\cite{Graham:2014sha} in order to calculate the sensitivity of a cylindrical receiver that is nested (or encapsulated) inside the emitter cavity~\cite{Kim:2020ask} (see e.g.~Fig.~\ref{fig:EncDesign} for detail) for sufficiently thin wall between receiver and emitter~\cite{Berlin:2023mti}. 

\begin{figure}[tbh!]\centering
\includegraphics[width=0.9\linewidth]{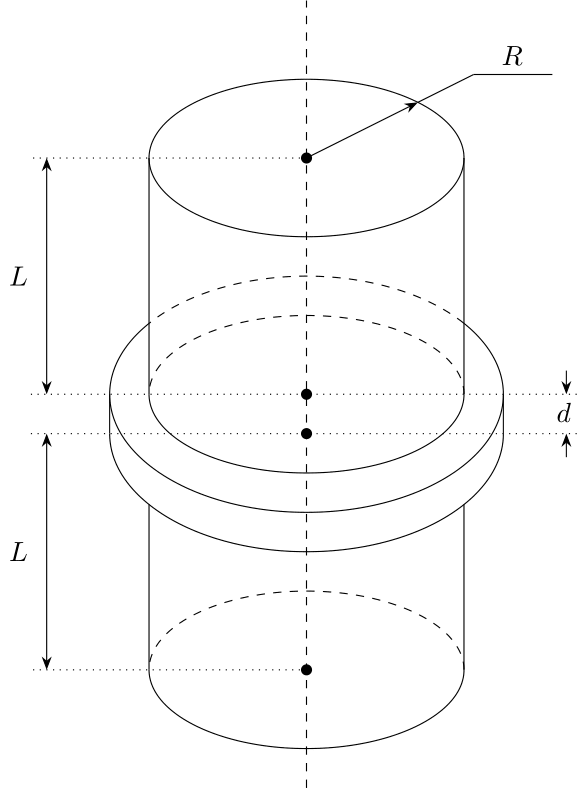}
\caption{A typical scheme of the setup for adjacent cavities. 
We note that a realistic configuration implies a thin conducting barrier between cavities such that $d \ll R,L$. }
\label{fig:DistDesign}
\end{figure}

The similar experimental setup is addressed in Ref.~\cite{Berlin:2023mti} as light-shining-through-thin-wall (LSthinW), but for adjacent cylindrical cavities (see e.g.~Fig.~\ref{fig:DistDesign} for detail).
 The typical barrier between the endcaps of the receiver and emitter cavity is considered to be significantly smaller than the characteristic dimensions of the cavities $d \ll 1/\omega$, where $\omega\sim \mathcal{O}(1)~\mbox{GHz}\sim~\mathcal{O}(1)~\mu\mbox{eV}$ is a typical driven radio frequency (RF) of the  emitter pump mode (it matches with the signal EM mode frequency that is resonantly induced in the hollow receiver). The minimum barrier thickness is determined by the requirement to suppress the emitter's electromagnetic field within the superconducting barrier, ensuring that it remains weaker than the level of expected signal as well as other types of noise in the receiver.
Numerically, we consider the minimal barrier thickness $d=10\,\mu$m which is $200$ times larger than the London penetration depth $\lambda_L \simeq 50 \, \mbox{nm}$ for superconducting Niobium, see~\cite{Berlin:2023mti}.

The  LSthinW setup allows enhancing the sensitivity in the off-shell DP region, i.e.~when hidden photon  mass, $\mA$, is relatively large $\mA \gg \omega$.
In addition, employing the sufficiently low driven frequency of the emitter, one can achieve a large density of the source photons and high quality of the superconducting RF (SRF) cavity~\cite{Berlin:2023mti}.   

In the present paper, we estimate the impact on the DP sensitivity that is associated with the wall thickness $d$ between cavities  for the nested design of LSthinW setup shown in~Fig.~\ref{fig:EncDesign}, that exploits both $\mbox{TE}_{011}$ and $\mbox{TM}_{010}$ pump modes.
We also show that there is an advantage of the nested receiver as opposed to its adjacent location Fig.~\ref{fig:DistDesign}  for the $\mbox{TM}_{010}$ pump mode and large mass approach, $m_{A'} \gg \omega$. Finally, we argue that for $\mbox{TE}_{011}$ mode,  both nested and adjacent  designs of the receiver  yield comparable sensitivities that are optimal to  the dark photon probing in  the large mass  region, $m_{A'} \gg \omega$.  
Moreover, the  expected reaches for both $\mbox{TM}_{010}$  and $\mbox{TE}_{011}$ modes can be  comparable for the nested receiver design in the large dark photon mass range, $m_{A'} \gg \omega$.

This paper is organized as follows. 
In Sec.~\ref{SectionFramework} we discuss general properties of the DP scenario and derive the equations of motion. 
In Sec.~\ref{SectionCavityResponce}  we  derive  a signal power that implies probing the DP in the SRF receiver. 
%In Sec.~\ref{Sectionadjacent} we calculate the general form-factor. 
In  Sec.~\ref{SectionResultsDSiscussion} we estimate the sensitivity of the aforementioned LSthinW designs.  
We conclude in Sec.~\ref{SecConclusion}. In the Appendix, we specify some useful formulas for the form-factor derivation in the analytical form. 

\section{Framework of the scenario
\label{SectionFramework}}

Let us consider the Lagrangian, describing a  coupling of the ordinary Standard Model (SM) photon, $\hat{A}_\mu$, with a massive vector hidden state, $\hat{A}_\mu'$, through the kinetic mixing 
\begin{equation}
    \mathcal{L} \supset - \frac{1}{4} \hat{F}_{\mu\nu}^2 - \frac{1}{4} \left(\hat{F}'_{\mu\nu}\right)^2 + \frac{1}{2} m_{A'}^2 \left( \hat{A}'_\mu \right)^2 + \frac{\epsilon}{2} \hat{F}_{\mu\nu}' \hat{F}^{\mu \nu } -  J_\mu \hat{A}^\mu, 
    \label{LagrFieldBasis}
\end{equation}
where we exploit the hat over the boson fields in order to label the states of the benchmark interaction Eq.~(\ref{LagrFieldBasis}) in so-called {\it field basis}~\cite{Graham:2014sha} or {\it kinetically mixed  basis}~\cite{Fedderke:2021aqo},
%$m_{A'}$ is a typical 
%mass of the dark photon field, 
$\epsilon$ is a dimensionless coupling of the kinetic mixing, that is assumed to be small, $\epsilon \ll 1$, $\hat{F}_{\mu\nu}^{(\prime)}$ is the stress tensor of the  (dark) photon, $J^\mu$ is the typical $U(1)$ electromagnetic  current density. 
Now let us get rid of the non-diagonal kinetic mixing term $\propto \epsilon $ in Eq.~(\ref{LagrFieldBasis}) by the following replacement of the field basis states $\hat{A}_{\mu}^{(\prime)}$ to the  {\it mass basis states}~$A_{\mu}^{(\prime)}$,
\begin{equation}
\hat{A}_\mu = A_\mu + \frac{\epsilon}{\sqrt{1-\epsilon^2}} A_\mu', \quad \hat{A}'_\mu = \frac{1}{\sqrt{1-\epsilon^2}} A'_\mu, 
\label{FielBasisdViaMassBasis}
\end{equation}
as a result, one gets the following redefined Lagrangian
\begin{align}
 \mathcal{L} \supset - \frac{1}{4} F_{\mu\nu}^2 - \frac{1}{4} \left(F'_{\mu\nu}\right)^2 + 
 \frac{1}{2} \frac{m_{A'}^2 }{(1-\epsilon^2)} \left(A'_\mu \right)^2  \nonumber \\
 - J_\mu \left(A^\mu + 
 \frac{\epsilon}{\sqrt{1-\epsilon^2}}A^{\prime \mu} \right),
 \label{MassBasisLagrangian1}
\end{align}
that leads to the decoupled equations of motion for both $A_\mu$ and $A_\mu'$
\begin{equation}
     \partial^2_\mu A_\nu = J_\nu,
     \label{MassBasisAEq1}
\end{equation}
\begin{equation}
    \partial^2_\mu A'_\nu +\frac{m_{A'}^2}{1-\epsilon^2} A_\nu' =\frac{\epsilon}{(1-\epsilon^2)} J_\nu,
    \label{MassBasisAprimeEq1}
\end{equation}
where we impose the Lorentz gauge $\partial_\mu A^\mu = 0$ and on-shell Proca condition $\partial_\mu A^{\prime \mu } = 0$, implying that it leads to the conservation of the electromagnetic current $\partial_\mu J^\mu=0$ (see e.g.~Ref.~\cite{Fedderke:2021aqo} for detail). The electric and magnetic fields are defined in a usual manner for both SM, $A_\mu$, and hidden photon, $A_\mu^\prime$,
\begin{equation}
    \E^{(\prime)} = - \bfnabla \phi^{(\prime)} -\partial_t \A^{(\prime)}, \qquad \B^{(\prime)} = \bfnabla \cdot\A^{(\prime)},
    \label{EandBdef1}
\end{equation}
where we imply the typical four-vector notation via potential, $A^{(\prime) 0} \equiv \phi^{(\prime)}$ and 
vector terms $ A^{(\prime) i } \equiv \A^{(\prime)}$. 
However, it was pointed out in Ref.~\cite{Graham:2014sha} that in order to calculate the physical state propagation one should express the fields in the mass basis through the linear combination of the  {\it visible},  $A^{\mu}_{\text{vis}}$,  and {\it invisible},  $A^{\mu}_{\text{inv}}$, modes  in the following form~\cite{Berlin:2023mti} 
\begin{align}
    A^{\mu}= \sqrt{1-\epsilon^2} A^\mu_{\text{vis} }  - \epsilon A^{\mu }_{\text{inv} } , 
    \\
    A^{ \prime \mu} =   \epsilon A^{ \mu }_{\text{vis} }  + \sqrt{1-\epsilon^2} A^{ \mu}_{\text{inv} }.  
\end{align}
As a result, for the regarding basis, the equations of motion (\ref{MassBasisAEq1}) and (\ref{MassBasisAprimeEq1}) can be rewritten as follows
\begin{align}
   \left(\partial^2_t - \bfnabla^2 \right)  A^\mu_{\text{vis}} = J^\mu_{\text{SM}} - 
   \frac{\epsilon m_{A'}^2}{(1-\epsilon^2)} A^{\prime \mu} \label{AvisViaApr}
    \\
     \left(\partial^2_t - \bfnabla^2  + \frac{ m_{A'}^2}{(1-\epsilon^2)}  \right) A^\mu_{\text{inv}} = 
    -  \frac{\epsilon m_{A'}^2}{(1-\epsilon^2)} A^{ \mu}, \label{AinvViaA}
\end{align}
where we use convention for the effective SM density current $J^\mu_{\text{SM}} \equiv \sqrt{1-\epsilon^2} J^\mu$. In Eqs.~(\ref{AvisViaApr}) and~(\ref{AinvViaA}) we keep the dependence on $A^{\prime \mu}$ and $A^\mu$ in the right-hand side respectively in order to emphasize that  at the leading order approach, $\epsilon \ll 1$, the visible component $A^\mu_\vis$  is sourced  by the electromagnetic SM current $J^\mu_{\text{SM}}$. 
Moreover, the invisible mode  $A^\mu_{\text{inv}}$ is sourced only by the typical effective current $\propto - \epsilon m_{A'}^2 A^{\mu}$, that is subdominant to the~$J^\mu_{\text{SM}}$. 

%\begin{figure}[tbh!]\centering
%\includegraphics[width=0.9\linewidth]{figures/Encapsulated_Design.pdf}
%\caption{Typical sketch of the cavity setup with the encapsulated receiver. %cavity.}
%\label{fig:EncDesign}
%\end{figure}

%\begin{figure}[tbh!]\centering
%\includegraphics[width=0.9\linewidth]{figures/adjacent_Design.pdf}
%\caption{Typical scheme of  the setup for  adjacent cavities. We note that a realistic configuration  implies thin conducting barrier between cavities such that $d \ll R,L$. }
%\label{fig:DistDesign}
%\end{figure}

\section{Cavity response
\label{SectionCavityResponce}}
Let us consider benchmark experimental setups consisting of a high-quality superconducting RF cylindrical receiver cavity, which is:
\begin{itemize}
\item nested within the emitter such that they share a common perfectly conducting thin side wall of thickness~$d \ll R$ 
 (see Fig.~\ref{fig:EncDesign});
\item adjacent to the emitter's end-cap, separated by a thin barrier of thickness $d \ll L$, where $L=L_{\emt}=L_{\rec}$ is the typical length of the cavities (see Fig.~\ref{fig:DistDesign}).
\end{itemize}

Let  $\xvec$ and $\yvec$ be the  coordinates associated with  the emitter and receiver frames, respectively. 
In addition, let $\lvec$ be the vector linking the origins of the emitter and receiver frames in the coaxial design of the experimental setup. In this case, there is a simple relationship between them: $\xvec = \lvec + \yvec$.  

The first cavity is pumped by a single electromagnetic mode (EM) 
\begin{eqnarray}
\E (\xvec,t) = \E_{\emt}(\xvec) e^{-i \omega t}, \label{EemDef1} \\
\B (\xvec,t) = \B_{\emt}(\xvec) e^{-i \omega t}, \label{BemDef1}
\end{eqnarray}
oscillating at frequency $\omega$, and serves as an emitter of dark photons.
In (\ref{EemDef1}) and (\ref{BemDef1}), we follow the convention of~\cite{Hill2014} for the time dependence~$\propto e^{-i \omega t}$ in order to match the notation of cavity eigenmodes~\cite{Hill2014}. 
The second cavity is a quiet hollow receiver of dark photons. 
Thus, the detecting EM mode is expected to be resonantly excited due to regeneration of a hidden vector state radiated from the emitter cavity.

Now  let us define the invisible electric field 
$\E_{\inv} \equiv \E' - \epsilon \E$, this means that Eq.~(\ref{AinvViaA}) implies 
the following massive wave equation
\begin{equation}
    \left( \partial^2_t - \bfnabla^2 + m_{A'}^2\right) \E_\inv  = - \epsilon m_{A'}^2 \E.
    \label{EqEinvViaE}
\end{equation}
The Eq.~(\ref{EqEinvViaE}) means that the invisible electric field inside the receiver  cavity sourced by the  massless field of the emitter $ \E = \E_\emt e^{-i \omega t}$. At the order of $\mathcal{O}(\epsilon)$, one has $\E_\inv \simeq \E'$, and thus the dark photon electric mode in the receiver can be expressed using the Green's function in the following form~\cite{Graham:2014sha}:
\begin{equation}
     \E'(\xvec, t ) = -\epsilon m_{A'}^2 \! \int\limits_{ V_{\emt}}
    \! d^3 \xvec' \, \frac{e^{-i \omega t +  i \kA |\xvec - \xvec'|}}{4\pi|\xvec - \xvec'|}  \E_\emt (\xvec'),
    \label{EprimeViaEem1}
\end{equation}
where the integration is performed over the volume of the emitter cavity in the emitter frame, $\kA$ is a wavenumber given by $\kA \equiv \sqrt{\omega^2 - m_{A'}^2}$ for relatively small masses, $m_{A'} \lesssim \omega$, and $\kA = i \varkappa_{A'}$, where $\varkappa_{A'}$ is defined as $\varkappa_{A'} \equiv \sqrt{m_{A'}^2 - \omega^2}$ for a heavy hidden vector state, $m_{A'} \gtrsim \omega$.

Now, let us estimate the resonant response of the receiver cavity. First, we define the visible electric field as $\E_\vis \equiv \E + \epsilon \E'$, which obeys the following boundary condition on the conducting surface:
\begin{equation} 
\left[ \nvec \cdot\E_\vis \right] \big|_{\partial V}=0, 
\label{GeneralBoundConcEn}
\end{equation}
where the unit vector $\nvec$ is defined along the normal to the interior boundary $\partial V $ of the cavity~\cite{Graham:2014sha}. 
One can show~\cite{Graham:2014sha} that, at leading order $\mathcal{O}(\epsilon)$, the receiver response $\E_\vis$ is sourced by the time derivative of the typical effective current:
\begin{equation}
\bfnabla\times(\bfnabla\cdot\E_\vis) + \partial^2_t \E_\vis  = -\partial_t \J_{\eff} (\yvec, t)
\label{EvisViaEprime1}, 
\end{equation}
where the left-hand side of the equation imply the derivatives with respect to the receiver frame coordinate $\yvec$,
%in what follows we shift the coordinates in the right hand side of Eq.~(\ref{EvisViaEprime1}), 
$\partial_t \J_{\eff} (\xvec, t) = -i \omega \J_{\eff} (\xvec) e^{-i \omega t}$, and the stationary part of the current is given by:
\begin{equation}
\J_{\eff} (\xvec) =  \frac{\epsilon}{i \omega} \left[ \bfnabla \cdot \left(\bfnabla \cdot \E' (\xvec) \right)-m_{A'}^2 \E' (\xvec)  \right],  
\end{equation}
where $\E'(\xvec)$ is a spatial profile term of Eq.~(\ref{EprimeViaEem1}). 
The solution to Eq.~(\ref{EvisViaEprime1}) for the receiver can be expressed in the following form:
\begin{eqnarray}
 \E_\vis (\yvec, t) & \simeq & -  \frac{Q}{\omega_n} \E^{(n)}_{\rec} (\yvec)  e^{-i \omega t } 
  \nonumber
\\
& \cdot &\!\! \frac{\int_{V_\rec} \! d^3 \! \yvec' \left(\E^{(n)*}_\rec (\yvec' ) 
  \cdot \J_{\eff}(\yvec')\right)}{\int_{V_{\rec}} d^3 \yvec' | \E_\rec^{(n)}(\yvec' ) |^2 },    
  \label{mainEvisReceiverResponse1}    
\end{eqnarray}
where $\E^{(n)}_\rec (\yvec)$ is a divergence-free vacuum cavitymode, 
$\bfnabla \cdot \E^{(n)}_\rec (\yvec)= 0$,  $Q$ is a quality factor of the receiver cavity, $\omega_n$ is a cavity eigenfrequency, which is close to the driving frequency $\omega_n  \simeq \omega$ in the resonant response regime, i.e.~when the signal builds up time $t$ is sufficiently large $ t\gg Q/\omega$. 
The quality factor is chosen to be $Q\simeq 10^{10}$ throughout the paper.  
For completeness, we note that the integration over $d^3 \yvec'$ in Eq.~(\ref{mainEvisReceiverResponse1}) is performed over the receiver cavity volume~$V_{\rec}$ in the receiver frame~$\yvec'$.

It can be shown~\cite{Berlin:2021txa} that the total energy stored in the receiver cavity due to the resonant response is given by:
\begin{equation}
     W_{\rec} \simeq \frac{1}{2} \int\limits_{V_\rec} d^3 \yvec \left[ \langle \mbox{Re}  [ \B_\vis (\yvec, t) ]^2 \rangle_t+   \langle   \mbox{Re} [ \E_\vis (\yvec, t) ]^2  \rangle_t \right]
    \label{Wrec1}
\end{equation}
where we employ the time averaging notations $\langle ... \rangle_t$ for the energy density of both the magnetic field, $\propto \langle   \mbox{Re} [ \B_\vis (\yvec, t) ]^2 \rangle_t /2 $, and the electric field, $\propto \langle   \mbox{Re} [ \E_\vis (\yvec, t) ]^2  \rangle_t/2$, in a perfect conductor. Here $\mbox{Re} [...]$ denotes the real part of the EM fields.
Remarkably, the energy densities are equal in the resonant regime~\cite{Berlin:2021txa}, and the resulting expression for $W_\rec$  can be expressed only through the integral over the electric energy density term, $\propto |\E_\vis (\yvec)|^2/2 $, where $\E_\vis (\yvec)$ is a stationary part of Eq.~(\ref{mainEvisReceiverResponse1}). 

From the discussion above, one can estimate the signal power emission of the cavity as
\begin{equation}
P_{\sign} \simeq \frac{\omega}{Q} W_{\rm rec} = \frac{\omega}{2Q} \int\limits_{V_\rec} d^3 \yvec |\E_{\vis}(\yvec)|^2 = \frac{Q}{2 \omega} \big| \mathcal{G} \big|^2,
\label{PsignDef1}
\end{equation}
where $\mathcal{G}$ is an overlapping factor (here we keep it dimensional for brevity), defined as
\begin{equation}
\mathcal{G}= \frac{\int_{V_\rec} d^3 \yvec \left(\E^{(n)*}_\rec (\yvec) 
\cdot \J_{\eff}(\yvec)\right)}{\left(\int_{V_{\rec}} d^3 \yvec | \E_\rec^{(n)}(\yvec ) |^2 )\right)^{1/2}}.
\label{GeomFactorDef1}
\end{equation}
It is worth noting Eq.~(\ref{GeomFactorDef1}) implies the signal is agnostic to overall constant normalization pre-factor of the receiver mode, $\E_\rec^{(n)}(\yvec )$; however, it depends on the specific mode type and the overlapping integrals in both the numerator and denominator of~Eq.~(\ref{GeomFactorDef1}). 

We estimate the sensitivity as the maximum output of the receiver, which is defined by the Dicke radiometer equation~\cite{Dicke:1946glx}:
\begin{equation}\label{Dicke}
\mbox{SNR} = \frac{P_\sign}{P_{\text{noise}}} \sqrt{ t \Delta \nu }, 
\end{equation}
where $t$ is an integration time of the signal, $\Delta \nu$ is a signal bandwidth, $P_{\text{noise}}$ is a power of noise. 

We assume the background consists of thermal noise as well as non-thermal noise from the radio frequency readout chain, as discussed in \cite{Kim:2020kfo}. Both noise components can be described by the Dicke radiometer equation as noise with an effective temperature $T_{sys} > T$, see \cite{Kim:2020kfo}. 
Thus, the noise power can be estimated as $P_{\text{noise}}\simeq T_{\rm sys} \Delta \nu $ for $\omega \ll T_{\rm sys}$. 
The typical effective temperature is chosen to be $T_{\rm sys}\simeq 4~\mbox{K}$. It is worth noting that the narrowest possible bandwidth can be as small as $\Delta \nu \simeq 1/t$~\cite{Bogorad:2019pbu}. 
In the numerical estimation, we conservatively set the integration time to be $t \simeq 1~\mbox{day} \simeq 8.6\cdot10^{4}~\mbox{s}$.

%{\bf \color{red} (delete?)
%We emphasize that niobium-based cavities can be utilized with a typical London penetration depth of approximately $\lambda \simeq 40 \, \mbox{nm}$. This suggests, in a conservative estimate, that the typical barrier thickness can be as large as $d \gtrsim 30 \, \lambda \simeq 1.2 ~\mu\mbox{m}$ \cite{Berlin:2023mti}. Therefore, our benchmark value of $d \simeq 10~\mu\mbox{m}$ represents a reasonable choice for wall thickness.
%}

%{\bf \color{blue} (This Discussion to Sec Conclusion?)
%Let us present an argument in support that the thin superconducting wall of $10\ \mu$m  width does not miss any electromagnetic field leakage from the emitter electromagnetic field to the receiver. Electromagnetic field strength decreases exponentially with the depth $x$ into the superconductor wall as $B = B_{em} e^{-x/\lambda}$, where $\lambda$ is the  London penetration depth,  which is $\lambda \simeq 50 \, \mbox{nm}$ for Niobium superconductor \cite{Berlin:2023mti}. The  minimal width of the wall necessary to suppress the emitter field is $x = \lambda \times \log(E_{em}/E)$. Taking  $E_{em} = 30\ \mbox{MV/m}$ and $E \sim 3\ \mu \mbox{V/m}$, one obtains $x \gtrsim 1.5\ \mu$m. Therefore, our benchmark value of $d \simeq 10~\mu\mbox{m}$ represents a reasonable choice for wall thickness: induced  electromagnetic mode  is significantly suppressed in the receiver.}

Let us present the general formula for calculating the sensitivity for the mixing parameter $\epsilon$ for the specific cavity geometry:
\begin{equation}
\epsilon = \left[ \frac{2T\cdot \mbox{SNR}}{ Q \omega^3 (E^0_{\rm em})^2 V^2_{\rm em} V_{\rm rec} t}\right]^{\frac{1}{4}}  \times m_{A'}^{-1}  |\widetilde{{\cal G}}|^{-1/2}.
\label{TypicalEpsilon1}
\end{equation}
where $\widetilde{{\cal G}}$ is a dimensionless geometric form factor, defined as
\begin{align}
\widetilde{{\cal G}} = & \dfrac{1}{\omega^3}  \int\limits_{V_{\rm rec}}  \dfrac{d^3 \yvec }{V_{\rm rec}} \int\limits_{V_{\rm em}} \dfrac{d^3 \xvec }{V_{\rm em}} \, {\cal E}_{\rm rec}^{*i}(\yvec )\, \cdot\label{defG}
\\ 
& \cdot(m^2_{A'}\delta_{ij} - \partial_{  i} \partial_{  j}) {\cal E}_{\rm em}^j(\xvec )  
\cdot \dfrac{\exp(ik_{A'}|\xvec - \yvec - \lvec|)}{4\pi|\xvec - \yvec - \lvec|}. \notag
\end{align}
Here, ${\cal E}^i_{\rec} (x )$  and  ${\cal E}^i_{\rec} (x')$  are cavity modes of the receiver and emitter respectively which are related to the electric fields as follows:
\begin{align}
 E_{\emt}^i (\xvec) = E^0_{\rm em} \cdot {\cal E}_{\rm em}^i (\xvec),
 \\
 E_{\rec}^i (\yvec) =   {\cal E}_{\rec}^i (\yvec), \label{calEzRecTM010}
\end{align}
where $E_0^\emt$ is the magnitude of the driven emitter electric field, indices $i$ and $j$ label the $i^{\text{th}}$ and $j^{\text{th}}$ components of the electric field respectively. 
The explicit expressions for ${\cal E}^i_{ \rec/\emt} ({\bf r})$ provided in~Appendix~\ref{AppendixEigenmodes}. 
In Eq.~(\ref{calEzRecTM010}), we omit the amplitude of the receiver field for brevity (see e.g.~discussion after Eq.~(\ref{GeomFactorDef1}) for details).
%We  introduce the pre-factor $1/\omega^3$ in Eq.~(\ref{defG}) in order to obtain dimensionless 
%form-factor $\widetilde{\cal G}$. We emphasize that notation 
%$a^i b_i$ represents the scalar product of two vectors: $(\vec{a} \cdot \vec{b})$, and the 
%symbol $\partial_i$ corresponds to the operator of nabla $\vec \nabla_{\vec{x}}$ with respect 
%to the coordinate $\vec x$. It is worth noting that the scalar product 
%$\partial_j {\cal E}^j_2(x')$ is calculated for vectors at two different points, $\vec{x}$ and 
%$\vec{x}'$. Therefore, in the case of a polar coordinate system, the result  contains an 
%additional factor associated with the difference in the bases.

The six-dimensional integral in the Eq.~(\ref{defG}) reduces to a one-dimensional integral, which can be calculated numerically in the general case and analytically in the limit of large dark photon masses, specifically for $m_{A'}\gg\omega$.
We employed the Fourier representation for the three-dimensional Green's function of the Helmholtz equation:
\begin{align}
    {\rm G}(\xvec - \yvec - \lvec) &= \dfrac{\exp(ik'|\xvec - \yvec - \lvec|)}{4\pi|\xvec - \yvec -\lvec|} 
\nonumber    
    \\
    &= \int \dfrac{d^3 k}{(2\pi)^3} \dfrac{e^{-i{\bf k}(\xvec - \yvec - \lvec)}}{|{\bf  k}|^2 - (k' + i\varepsilon)^2},
    \label{Helmholz}
\end{align}
where $i\varepsilon$ is an infinitesimally small imaginary term added to pick up poles on the complex plane. The details of form-factor calculation are provided in Appendix~\ref{HelpfulIntegrals}. 
In addition, we note that eigenmodes are normalized according to the following condition:
\begin{equation} 
\int\limits_{V_{\rec/\emt}} \! d^3 {\bf x}  \ |{\cal E}^i_{ \rec/\emt} ({\bf x})   |^2 = V_{\rec/\emt}.
\label{NormCondDimslsE}
\end{equation}

In the present paper, we set the typical driving EM field to be $E_0^\emt= 30~\mbox{MV}/\mbox{m}$, $B_0^\emt=0.1~\mbox{T}$  (see e.g.~Ref.~\cite{Berlin:2023mti} for detail). 
In order to estimate the expected reach of the experimental configuration, we put $\mbox{SNR}\simeq 5$ in Eq.~(\ref{TypicalEpsilon1}).  
%$ E^0_{em}$ is the amplitude for the field in the emitter,
%${\cal E}_{\rm rec}^i =  E_{\rm rec}^i /  ||E_{\rm rec}|| $,  
%$ E_{\rm em}^i  = E^0_{\rm em} \cdot {\cal E}_{\rm em}^i$.

 The assumption of a small barrier thickness requires consideration of potential electromagnetic field leakage from the emitter to the receiver. Since the electromagnetic field strength decreases exponentially with depth $x$ into the superconductor wall, given by $B(x) = B_{\rm em} e^{-x/\lambda_L}$, where $\lambda_L$ is the  London penetration depth, the barrier thickness must be sufficient to ensure that the power of the attenuated emitter field is significantly lower than both the signal and the noise level.
\cite{Berlin:2023mti}.

The sensitivity level for the signal magnetic field in the receiver cavity can be derived from the Eqs.~(\ref{PsignDef1}),~(\ref{Dicke}):
\begin{equation}
    B_{\rm sig} = \left[\dfrac{2 T_{\rm sys} Q\, {\rm SNR}}{V\omega  t}\right]^{\frac{1}{2}}. 
\end{equation}
Taking the typical values of the parameters:
%ЕFor the typical parameters,
%\begin{equation}
$    {\rm SNR = 5}, \ T_{\rm sys} = 4\, {\rm K}, \  t = 1 \, {\rm day},$ 
%\end{equation}
%\begin{equation}
$R = 10 \, {\rm cm}, \ L = 40  \, {\rm cm},$
%\end{equation}
one obtains the characteristic ratio of signal magnetic field sensitivity in the receiver cavity, and the pump magnetic field $B_{\rm pump} = 0.1 \,$T in the emitter cavity: %the  and minimal required thickness are
\begin{equation}
    \dfrac{B_{\rm sig}}{B_{\rm pump}} \simeq 10^{-14},
\end{equation}
and the minimal required barrier thickness, given by:
\begin{equation}
    d_{\rm min} = -\lambda_L \ln \dfrac{B_{\rm sig}}{B_{\rm pump}} \simeq 30 \lambda_L = 1.5 \, \mu\mbox{m}. 
\end{equation}
The chosen value of $d = 10\,\mu$m is nearly $7$ times larger than $d_{\rm min}$, so the emitter magnetic field does not pass through $10\,\mu$m barrier with a guarantee.

%Taking characteristic values for physical parameters and assuming $\Delta \nu \simeq 1/t$ and $\mbox{SNR} \simeq 5$ one obtains,
%\begin{align}
%\epsilon =4.9\cdot 10^{-11} \frac{10^{-10} \rm eV}{m_{A'}}\left(\frac{4\cdot 10^{10}}{Q}\right)^{\frac{1}{4}}\left(\frac{1 \rm m^3}{V_{\rm em}}\right)^{\frac{1}{2}} \left(\frac{1 \rm m^3}{V_{\rm rec}}\right)^{\frac{1}{4}} \notag \\ \left(\frac{T}{4\,\rm K}\right)^{\frac{1}{4}}  \left(\frac{1\,\rm GHz^3}{\omega^3}\right)^{\frac{1}{4}}  \left(\frac{0.1\, T}{E^0_{\rm em}}\right)^{\frac{1}{2}} \left(\frac{1 \rm year}{t}\right)^{\frac{1}{4}}  |\widetilde{{\cal G}}|^{-1/2}.
%\end{align}

%\color{blue}
%\section{Geometric form factor calculation}

%Let us consider the reduction of the geometric form factor calculation. We define the dimensionless form factor $\tilde{\cal G}$ as follows,
%\begin{widetext}
%\begin{equation}\label{defG}
%    \tilde{\cal G} \stackrel{\rm def}{\equiv} \dfrac{1}{\omega^3} \int_{V_{\rm rec}} \dfrac{d^3 x}{V_{\rm rec}} \int_{V_{\rm em}} \dfrac{d^3 x'}{V_{\emt}} \, {\cal E}^{*\, i}_{\rm rec}(x) \left[m^2_{A'}\delta_{ij} - \partial_i\partial_j\right]  {\cal E}_\emt^j(x') \times \left\{ -\dfrac{\exp(ik'|\vec{x} - \vec{x}' - \vec{l}|)}{4\pi|\vec{x} - \vec{x}' -\vec{l}|} \right\},
%\end{equation}
%\end{widetext}

%\section{Results and discussions}

%\vspace{1.5cm}

\section{Results and discussion} 
\label{SectionResultsDSiscussion}

%\begin{figure}[h!]\centering
%\includegraphics[width=0.9\linewidth]{figures/adjacent.png}
%\includegraphics[width=0.9\linewidth]{figures/encapsulted.png}
%\caption{The setup sensitivity to dark photon parameters for TM$_{010}$ mode. 
%Top panel: adjacent configuration, bottom panel: encapsulated configuration. The solid and dashed lines represent the typical parameters $R = R_3 = 5 \, {\rm cm}, L = 20 \, {\rm cm}$ and  $R = R_3 = 10 \, {\rm cm}, L = 40 \, {\rm cm}$, respectively. Green lines correspond to the typical barrier thickness of $ d= 1000~\mu\mbox{m}$, blue lines are associated with $d= 400~\mu\mbox{m}$, and red lines show the ultimate limits for $d= 10~\mu\mbox{m}$.  We set the typical setup parameters to be  $Q=4\cdot10^{10}$, $E^0_\emt = 30 \mbox{MV}/\mbox{m}$, $T=4~\mbox{K}$, and $t\simeq 1~\mbox{year}$. We also show current experimental regions that have been already ruled out by DarkSRF Pathfinder~\cite{Romanenko:2023irv}, CROWS~\cite{Betz:2013dza},  and XENON1T~\cite{An:2020bxd}, these bounds were adapted from Ref.~\cite{Berlin:2023mti}}
%\label{fig:1}
%\end{figure}

For the calculation of geometrical form factor at the large masses, we developed an independent analytical method, compared to the approach in~\cite{Berlin:2023mti}, and obtained the asymptotic sensitivity including exact pre-factors for $\epsilon$ (see e.g.~Appendix.~\ref{SecAsympt}) and typical mass dependencies (see e.g.~Tab.~\ref{tab:resultsMass}). 
The dependencies of sensitivity on the aspect ratio are presented in the Tab.~\ref{tab:results}. 
In the case of the adjacent cavity design and the TE$_{011}$ pump mode, we  independently reproduce the explicit result of~\cite{Berlin:2023mti} for the form factor. 

Remarkably, the $\epsilon \propto (R/L)^{-1/2}$ scaling for the adjacent case of $\mbox{TE}_{010}$ mode implies that the optimal cavity design is associated with a {\it  pancake-like} geometry, where $R \gg L$, with a large radius and small length (as also pointed out in  Ref.~\cite{Berlin:2023mti}).
Moreover, the sensitivity in this case is agnostic to the aspect ratio scaling, $R\to \alpha R$ and $L \to \alpha L$, where $\alpha$ is a positive scaling parameter.
The latter statement also holds for the nested geometry for both $\mbox{TE}_{011}$ and $\mbox{TM}_{010}$ pump modes, due to the aspect ratio scaling $\epsilon \propto (R_1/L)^{1/4}$. 
Nested design implies that the optimal geometry of the setup would be a {\it cigarette-like}  cavity with $L \gg R_1$. This is in contrast to the adjacent design of $\mbox{TM}_{010}$ mode, for which DP coupling scales as $ \epsilon \propto (RL)^{1/4}$. 
For the latter case, one can achieve an enhancement of the sensitivity in the large mass region, $m_{A'} \gg \omega$, by reducing the aspect ratio, $ R \to \alpha R $ and $ L \to \alpha L$, for $ \alpha \ll 1$. 
In what follows, we consider benchmark {\it pancake-like } geometry ($R = 20 \, {\rm cm}, L = 5 \, {\rm cm}$) for adjacent design and {\it cigarette-like} configuration for nested design ($ R_3 \simeq 10 \, {\rm cm}, L = 40 \, {\rm cm}$). 

{\setlength{\tabcolsep}{0.5em}
\renewcommand{\arraystretch}{2}% for the vertical padding
\begin{table}[!ht]
 %   \color{blue}
    \centering
    \begin{tabular}{ c c c c }
    \hline
    \hline
    Geometry & Mode  &  $m_{A'} \ll \omega$ &  $m_{A'} \gg \omega$ \\
    \hline
    \multirow{2}{*}{Adjacent} & TM$_{010}$ & 
    %\cellcolor{green!10}
    $m_{A'}^{-1}$  & 
    %\cellcolor{red!10} 
    $m_{A'}^{3/4} \cdot \exp(m_{A'} d / 2)$ \\
    \cline{2-4}
         &  TE$_{011}$ & 
         %\cellcolor{red!10} 
         $m_{A'}^{-2}$  & 
         %\cellcolor{green!10} 
         $m_{A'}^{1/2} \cdot \exp(m_{A'} d / 2)$  \\
    \hline
    \multirow{2}{*}{Nested} & TM$_{010}$ & 
    %\cellcolor{yellow!10} 
    $m_{A'}^{-1}$ & 
    %\cellcolor{green!10}
    $m_{A'}^{1/2} \cdot \exp(m_{A'} d / 2)$  \\
    \cline{2-4}
    &  TE$_{011}$ & 
    %\cellcolor{red!10} 
    $m_{A'}^{-2}$ & 
    %\cellcolor{green!10}
    $m_{A'}^{1/2} \cdot \exp(m_{A'} d / 2)$  \\
    \hline
    \hline
    \end{tabular}
    \caption{Asymptotic behaviour of $\epsilon$ at low and large masses for both adjacent and nested geometries and two types of cavity pump modes.}
    \label{tab:resultsMass}
\end{table}
}

{\setlength{\tabcolsep}{0.5em}
\renewcommand{\arraystretch}{2}% for the vertical padding
\begin{table}[!ht]
 %   \color{blue}
    \centering
    \begin{tabular}{ c c c }
    \hline
    \hline
    Geometry & Mode  &  $m_{A'} \gg \omega$ \\
    \hline
    \multirow{2}{*}{Adjacent} & TM$_{010}$ & 
    %\cellcolor{yellow!10} 
    $\epsilon \propto (RL)^{1/4}$   \\
    \cline{2-3}
         &  TE$_{011}$  & 
         %\cellcolor{green!10} 
         $\epsilon   \propto \left(R/L\right)^{-1/2}$  \\
    \hline
    \multirow{2}{*}{Nested} & TM$_{010}$  &  
    %\cellcolor{red!10} 
    $\epsilon  \propto \left(R_1/L\right)^{1/4}$  \\
    \cline{2-3}
    &  TE$_{011}$  &  
    %\cellcolor{red!10}
    $\epsilon  \propto \left(R_1/L\right)^{1/4}$  \\
    \hline
    \hline
    \end{tabular}
    \caption{Geometrical prefactor of $\epsilon$ asymptotic behaviour at large masses for both adjacent and nested geometries and two types of cavity pump modes.}
    \label{tab:results}
\end{table}
}

\begin{figure}[h!]\centering
\includegraphics[width=0.9\linewidth]{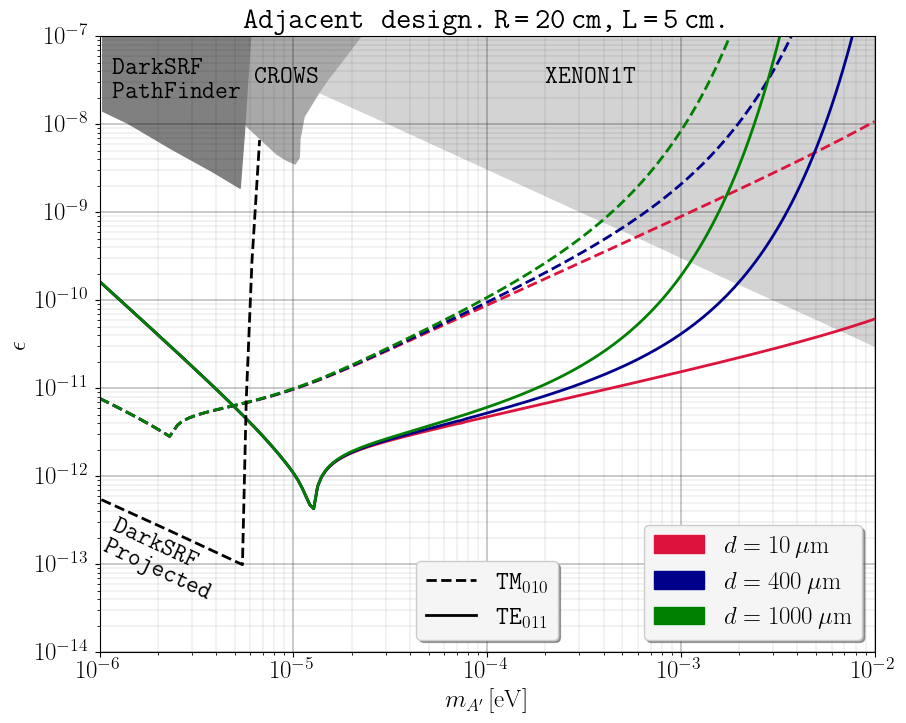}
\includegraphics[width=0.9\linewidth]{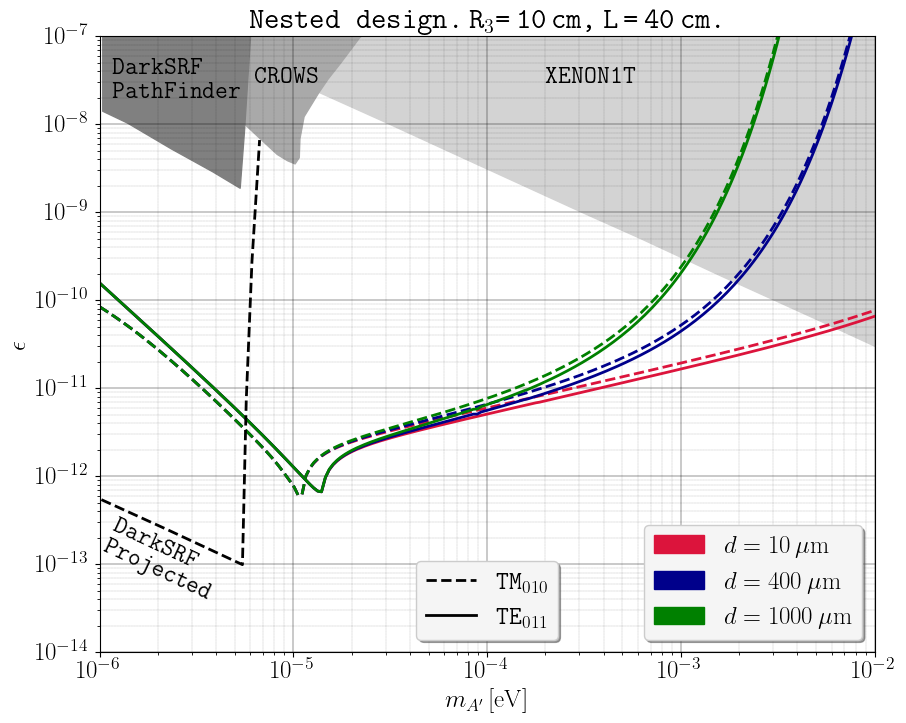}
\caption{The sensitivity to dark photon parameters for the TM$_{010}$ and TE$_{011}$ modes. 
Top panel: adjacent configuration, bottom panel:  nested configuration. %
The solid lines represent the TE$_{011}$ mode and the dashed lines represent the TM$_{010}$ mode. Typical parameters are $R = 20 \, {\rm cm}, L = 5 \, {\rm cm}$ for the adjacent configuration, and $R_3 = 10 \, {\rm cm}, L = 40 \, {\rm cm}$ for the nested configuration. The green lines correspond to the typical barrier thickness of $ d= 1000~\mu\mbox{m}$, the blue lines are associated with $d= 400~\mu\mbox{m}$, and the red lines show the ultimate limits for $d= 10~\mu\mbox{m}$.  
We set the typical setup parameters as follows:  
$Q=10^{10}$, $E^0_\emt = 30 \, \mbox{MV}/\mbox{m}$ $(B^0_\emt = 0.1\,\mbox{T})$, $T=4~\mbox{K}$, and $t = 1 \, \mbox{day} \simeq 8.6\cdot10^{4}~\mbox{s}$.
We also indicate current experimental regions that have been already ruled out by DarkSRF Pathfinder~\cite{Romanenko:2023irv}, CROWS~\cite{Betz:2013dza},  and XENON1T~\cite{An:2020bxd}. These bounds were adapted from Ref.~\cite{Berlin:2023mti}}
\label{fig:1}
\end{figure}

In  Fig.~\ref{fig:1} we show the sensitivities to the dark photon coupling $\epsilon$ and mass $m_{A'}$, for two different experimental configurations: adjacent  receiver (top panel) and encapsulated receiver (bottom panel). 
The expected reaches are shown for the barrier thickness in the range $10~\mu\mbox{m} \lesssim d \lesssim 1~\mbox{mm}$. 
The code for numerical calculations and figures generation can be found here \footnote{\href{https://github.com/salnikov-dmitry/lsw}{https://github.com/salnikov-dmitry/lsw}}.

Given the benchmark geometric configuration ($R \simeq 20~\mbox{cm}$, $L\simeq 5~\mbox{cm}$) for the adjacent setup, the resonant enhancement of the DP sensitivity $\epsilon \lesssim 4\cdot 10^{-13}$ and $\epsilon \lesssim 2\cdot10^{-12}$ can be achieved for the $\mbox{TE}_{011}$ and $\mbox{TM}_{010}$ modes respectively. 
The expected reach for off-shell regime, $\mA \gg \omega$, for $\mbox{TM}_{010}$ can be as small as $\epsilon \lesssim 10^{-9}$ for the masses below $\mA \lesssim  10^{-3}~\mbox{eV}$. However, for the $\mbox{TE}_{011}$ mode, one would achieve the ultimate sensitivity at the level of  $\epsilon \lesssim  2 \cdot 10^{-11}$ for $\mA \lesssim  10^{-2}~\mbox{eV}$.

The nested design of the emitter for the radius $R_3 \simeq 10~\mbox{cm}$ and the typical length $L \simeq 40~\mbox{cm}$ can provide the peak sensitivity at the level of $\epsilon \lesssim 6\cdot 10^{-13}$ for resonant mass $\mA \lesssim 10^{-5}~\mbox{eV}$ for both $\mbox{TE}_{011} $ and $\mbox{TM}_{010}$ modes.  
The ultimate thickness of the barrier between the emitter and the receiver, $d\simeq 10~\mu\mbox{m}$, can ensure the off-shell DP, $\mA \gg \omega$, sensitivity at the level of $\epsilon \lesssim 2\cdot 10^{-11}$ for the mass range below $\mA \lesssim 10^{-2}~\mbox{eV}$.

It is shown in Fig.~\ref{fig:1} that the LSthinW sensitivity bound in case of off-shell dark photons, $m_{A'} \gtrsim \omega$ can be associated with two typical regimes. 
The first one is associated with the mass range $m_{A'} \gtrsim d^{-1}$ that relates to the  exponential decrease of the sensitivity.
The second one is  the intermediate regime for $\omega \lesssim  m_{A'} \lesssim d^{-1}$. 
In this mass range, the sensitivity is still high, and the typical $\epsilon$ increases as a power law function of the dark photon mass. 

We reproduce the result from Ref.~\cite{Berlin:2023mti} for the adjacent cavity geometry, and conclude that the $\mbox{TE}{011}$ mode is more optimal for the dark photon search in the large mass regime, where $m{A'} \gg \omega$. 
However, for this geometry type, the TM$_{010}$ pump mode corresponds to the most  preferable option for the low mass limit, $m_{A'} \ll \omega$. The encapsulated geometry implies the similar dependencies of sensitivity on mass for both TM$_{010}$ and TE$_{011}$ modes. 
They exhibit the same asymptotic behaviour at large masses as the adjacent setup with the TE$_{011}$ mode; however, the sensitivity at low masses is weaker compared to the TM$_{010}$ mode in the adjacent geometry. 
 
% The slope of the curves in the encapsulated configuration is  flatter  than in 
% the  adjacent one.  Thus, we can conclude that the encapsulated receiver setup can 
% provide an enhanced sensitivity  for the case of large   
% dark photon  mass and the narrow wall between cavities. 

% The  possible extension of the present work can be addressed to a probing of axion-like 
%particles~\cite{Salnikov:2023haw,Gao:2020anb,Berlin:2022hfx} in LSthinW setup for both adjacent and 
%encapsulated receiver. We leave that task for future work.  

\section{Conclusion
\label{SecConclusion}}

% \cite{Berlin:2023mti} 
% \cite{Kim:2020ask}

We have considered the light-shining-through-thin-wall cavity setup for probing Dark Photons implying two experimental designs: (i) the receiver cavity is adjacent from the emitter's end-cap by the thin barrier (adjacent design), (ii)  the receiver is nested into the emitter cavity such that  they have a common thin side wall (nested design). 

The nested design was originally  suggested by the authors of Ref.~\cite{Kim:2020ask} for dark photon probing in case of  zero thickness of the side wall between the emitter and receiver. 
However, in the present paper we argue that finite thickness of the barrier between cavities can impact the dark photon sensitivity for the large mass region, $m_{A'} \gg \omega$. 
In addition, we explicitly calculated the dark photon sensitivities of the above-mentioned designs for both TM$_{010}$ and TE$_{011}$ pump modes of the receiver. 

We explicitly reproduce the result of Ref.~\cite{Berlin:2023mti} for the geometrical form factor, that implies the TE$_{011}$ pump mode and adjacent cavity design. 
We also show that there is an advantage of the nested receiver as opposed to its adjacent location for the  $\mbox{TM}_{010}$ pump mode and large mass approach, $m_{A'} \gg \omega$.

Finally, we argue that for the $\mbox{TE}_{011}$ mode, both nested and adjacent receiver designs yield comparable sensitivities, which can be optimal for the dark photon probing in the large mass region, $m_{A'} \gg \omega$.  
Furthermore, the expected reaches for both $\mbox{TM}_{010}$ and $\mbox{TE}_{011}$ modes can be comparable for the nested receiver design in the large dark photon mass region, $m_{A'} \gg \omega$. 
 
We derived explicitly the typical dependence of the expected reach on the aspect ratio of the cavity for the nested design, that yields $ \epsilon \propto (R_1/L)^{1/4}$. 
This is associated with the optimal {\it cigarette-like} design $(L \gg R_1)$ of the regarding setup.

%? We summarized the results of the work~\cite{Kim:2020ask, Berlin:2021txa} and developed the 
%remaining geometry cases. ?

\section{Acknowledgements}
This work  is supported by  RSF  grant no. 21-72-10151.  
%The work of L.V. on  numerical calculation of overlapping integrals is supported by the BASIS
%Foundation, grant no 23-2-9-30-1.

\appendix
\section{Eigenmodes
\label{AppendixEigenmodes}}

{\it TM$_{010}$ mode for  adjacent cavities: }
%\label{Sectionadjacent}}
 this mode implies 
the following  non-zero electric field components along $z$-direction
\begin{eqnarray}
   {\cal E}_{z, \emt}^{\text{TM}_{010}} (\rho,\phi,z)  =  J_0 \left(k_{1\rho} \rho \right)/J_1(\omega R), 
   \label{EzEmTM010}
   \\
   {\cal E}_{z, \rec}^{\text{TM}_{010}}  (\rho',\phi',z')   = J_0 \left(k_{1\rho} \rho' \right)/J_1(\omega R),
   \label{EzRecTM010}
\end{eqnarray}
where $(\rho,\phi,z)$ and $(\rho',\phi',z')$ are the cylindrical coordinates associated with emitter and 
receiver frame respectively, $k_{1\rho}$ is ratio $k_{1 \rho} =x_{01}/R$  with $x_{01}$ being a first zero 
of Bessel function,  $J_0 (x_{01})=0$. We take into account in Eqs.~(\ref{EzEmTM010}) 
and~(\ref{EzRecTM010}) that the tangential component of the electric field is zero on the conducting 
surface~(\ref{GeneralBoundConcEn}).

{\it TM$_{010}$ mode for nested cavity:}
% \label{SectionEncapsulated}}
we imply  that the thickness of the  superconducting wall 
between cavities~\cite{Berlin:2023mti} is sufficiently small 
$d \equiv R_2-R_1  \ll R_{3}\simeq \mathcal{O}(10)\mbox{cm} $.  
We note that Eqs.~(\ref{GeneralBoundConcEn}) and~(\ref{EzRecTM010})
imply that the electric field mode  of the  cylindrical receiver is  given  by 
\begin{equation}
       {\cal E}_{z, \rec}^{\text{TM}_{010}}  (\rho',\phi',z')   = 
       J_0 \left( \omega \rho' \right)/J_1(\omega R_1),
   \label{EzRecEncTM010}
\end{equation}
where $\omega = x_{01}/R_1$ is an eigenfrequency of TM$_{010}$ mode.
The emitter electric field can be expressed through the specific cylindrical
function $Z_{0}(x)$ in the following form,
\begin{eqnarray}
&  {\cal E}_{z, \emt}^{\text{TM}_{010}} &  \!\! (\rho, \phi, z)  
\label{EzEmtEncTM010}
 \\ 
& = &\beta\left(J_0(\omega \rho) - \dfrac{J_0(\omega R_2)}{Y_0(\omega R_2)} \cdot Y_0(\omega \rho)\right) \equiv \beta Z_0(\omega \rho), \nonumber
\end{eqnarray}
where $Y_0(x)$ is the Neumann function of zeroth order. The pre-factor $\beta$ in 
Eq.~(\ref{EzEmtEncTM010}) can be obtained
from the normalization integral Eq.~(\ref{NormCondDimslsE}), 
\begin{equation}
\beta = \left[\dfrac{R^2_3 -R^2_2}{R^2_3Z_1^2(\omega R_3) - R^2_2 Z_1^2(\omega R_2)}\right]^{1/2},
\label{BetaCoeffEnc}
\end{equation}
where $R_3$ is a solution to the equation $Z_0(\omega R_3)\equiv 0$, 
in  Eq.~(\ref{BetaCoeffEnc}) we also  exploit an additional   
cylindrical function in the following  form 
$Z_1(x) \equiv J_1(x) -  Y_1(x) J_0(\omega R_2)/Y_0(\omega R_2)$. 

{ \it  TE$_{011}$ mode for nested cavity:}
%\label{GfactorTE011NestedDesign}}
now we summarize main formulas for nested design of the cavities and TE$_{011}$ mode. 
Note that the non-zero electric field mode  of the  cylindrical receiver is  given  by
\begin{equation}
       {\cal E}_{\phi, \rec}^{\text{TE}_{011}}  (\rho',\phi',z')   = \alpha
       J_1 \left( k_{1\rho} \rho' \right) \sin \left( k_{1z} z' \right) ,
   \label{EzRecEncTE011}
\end{equation}
where $k_{1\rho} \equiv x_{11}/R_1$ and $k_{1z}\equiv\pi/L$ are the wave vectors for the 
$(n,p,q)=(0,1,1)$ quantum numbers respectively,
with $x_{11}$ being a first zero of the Bessel function $J_{1}(x_{11})=0$; the typical eigen-frequency of TE$_{011}$ mode is   $\omega =\left(k_{1\rho}^2 + k_{1z}^2 \right)^{1/2}$, the normalization factor reads
$\alpha =\sqrt{2}/|J_{2}(x_{11})|$. The electric field  of the layer-like emitter reads
\begin{eqnarray}
&  {\cal E}_{\phi, \emt}^{\text{TE}_{011}} &  \!\! (\rho, \phi, z)  \equiv \beta Z_1(k_{1\rho} \rho)  
\label{EzEmtEncTE011}
 \\ 
& = &\beta\left(J_1(k_{1\rho} \rho) - \dfrac{J_1(k_{1\rho} R_2)}{Y_1(k_{1\rho} R_2)} \cdot Y_1(k_{1\rho} \rho)\right). \nonumber
\end{eqnarray}
The pre-factor $\beta$ in  Eq.~(\ref{EzEmtEncTE011}) can be obtained
from the normalization integral Eq.~(\ref{NormCondDimslsE}), 
\begin{equation}
\beta = \left[\dfrac{2(R^2_3 -R^2_2)}{R^2_3 Z_2^2(k_{1\rho } R_3) - R^2_2 Z_2^2(k_{1\rho} R_2)}\right]^{1/2},
\label{BetaCoeffEncTE011}
\end{equation}
where $R_3$ is a solution to equation $Z_1(k_{1 \rho}R_3)=0$ and the auxiliary function reads
$Z_2(x) \equiv J_2(x) -  Y_2(x) J_1(k_{1 \rho} R_2)/Y_1( k_{1 \rho} R_2)$. Note that  the following identities hold $Z_2(k_{1 \rho}R_{3,2}) =-Z_0(k_{1 \rho}R_{3,2})$ due to 
$Z_1(k_{1 \rho}R_3) =Z_1(k_{1 \rho}R_2)=0$.

{\it TE$_{011}$ mode for adjacent cavities: }
%\label{GfactorTE011adjacentDesign}}
the non-zero electric field components are
\begin{eqnarray}
       {\cal E}_{\phi, \rec}^{\text{TE}_{011}}  (\rho',\phi',z')   = \alpha
       J_1 \left( k_{1\rho} \rho' \right) \sin \left( k_{1z} z' \right),
   \label{EphiRecSepTE011}
\\
       {\cal E}_{\phi, \emt}^{\text{TE}_{011}}  (\rho,\phi,z)   = \alpha
       J_1 \left( k_{1\rho} \rho \right) \sin \left( k_{1z} z \right),
   \label{EphiEmtSepTE011}
\end{eqnarray}
where $k_{1\rho} \equiv x_{11}/R$ and $k_{1z}\equiv\pi/L$ are the wave numbers for the 
$(n,p,q)=(0,1,1)$. 

\section{Auxiliary integrals
\label{HelpfulIntegrals}}
In this section, we summarize some helpful integrals for the calculation of 
the form-factor~(\ref{defG}), implying Eq.~(\ref{Helmholz}).   The momentum space is considered in the cylindrical coordinate system, $\vec{k} \to (k_\rho, \varphi_k, k_z)$, $d^3k = k_\rho dk_\rho d\varphi_k dk_z$.
Performing a change in the order of integration, we analytically calculated integrals over $x$, $x'$ and an angle in momentum space $\varphi_k$ in  all geometrical and pump modes cases. 

Integrating over spaces and momentum angles $\varphi, \varphi', \varphi_k$:
\begin{align}
    &\int_0^{2\pi}d\varphi \int_0^{2\pi}d\varphi' \int_0^{2\pi}d\varphi_k \, f_0(\vec{x};\vec{x}';\vec{k})   \\
    &=(2\pi)^3 \times J_n(k_\rho\rho)J_n(k_\rho\rho') \times f_1(\rho,z;\rho',z;k_\rho,k_z),
    \nonumber
\end{align}
where $n = 0$ in the case of TM$_{010}$ mode and $n=1$ in the case of TE$_{011}$ mode.

Integrating over $z$ and $z'$:
\begin{align}
    &\int_0^L dz \int_0^L dz' \, f_1(\rho,z;\rho',z;k_\rho,k_z)  \\
    &= L^2\times \left[ \dfrac{\sin(\frac{k_zL}{2})}{\frac{k_zL}{2}}\right]^2 \times f_2(\rho; \rho'; k_\rho, k_z), \nonumber
\end{align}
for the TM$_{010}$ mode and 
\begin{align}
    &\int_0^L dz \int_0^L dz' \, f_1(\rho,z;\rho',z;k_\rho,k_z) \\
    &= \left(\dfrac{L}{2}\right)^2\times \left[ \dfrac{ \pi \cos(\frac{k_zL}{2})}{\left(\frac{k_zL}{2}\right)^2 - \left(\frac{p_zL}{2}\right)^2} \right]^2 \times f_2(\rho; \rho'; k_\rho, k_z),
     \nonumber
\end{align}
for the TE$_{011}$ mode, $p_z = \pi/L$ is axial eigenmomentum.

Integrating over $\rho$ and $\rho'$:
\begin{align}
    & \int_{0}^{R} \rho d\rho  \, f_2(\rho;\rho';k_\rho,k_z)  \\
    &=R^2 \times \dfrac{p_\rho R J'_n(p_\rho R) J_n(k_\rho R)}{(k_\rho R)^2 - (p_\rho R)^2} \times f_3(\rho'; k_\rho, k_z), \nonumber
\end{align}
for the cylinder and
\begin{align}
    &\int_{R_2}^{R_3} \rho d\rho \, f_2(\rho;\rho';k_\rho,k_z)  \\
    &=R_1^2  \dfrac{p_\rho R_3  Z'_n(p_\rho R_3) J_n(k_\rho R_3) - \{3 \to 2\}} {(k_\rho R_1)^2 - (p_\rho R_1)^2}  f_3(\rho'; k_\rho, k_z), \nonumber
\end{align}
for the cylinder layer. Where $k_\rho = x_{n1}/R$ is radial eigenmomentum ($n = 0$ for TM$_{010}$ and $n=1$ for TE$_{011}$).

In the case of adjacent geometry, we performed integration for the cylinder twice (over both $\rho$ and $\rho'$). In the case of encapsulated geometry, we performed integration for cylinder over $\rho$ and for cylinder layer over $\rho'$.

The remaining two-dimensional integral in momentum space can be reduced by contour integration to a one-dimensional one according to Jordan's lemma. Namely, in the case of adjacent geometry ($l_z \neq 0$) by integration over $k_z$ variable, and in both adjacent and encapsulated geometries by integration over $k_\rho$.

Contour integration over $k_\rho$:
\begin{align}\label{k_rho_integration}
&\int^{+\infty}_{0} k_\rho dk_\rho \, \dfrac{J_n(k_\rho R')J_n(k_\rho R)}{[(k_\rho R)^2 - (p_\rho R)^2]^2} \cdot \dfrac{1}{k_\rho^2 - (q + i\varepsilon)^2} \nonumber  \\
&= \dfrac{\pi i}{2} \left[\dfrac{H^{(1)}_n(qR')J_n(qR)}{[(qR)^2 - (p_\rho R)^2]^2} - \dfrac{iY_n(p_\rho R')J'_n(p_\rho R)}{2(p_\rho R)[(qR)^2 - (p_\rho R)^2]} \right]  \nonumber \\
& =\dfrac{\pi i}{2} \left[ F(q) - \tilde{F}(q) \right],
\end{align}
where $R' \geq R$ and $q^2 = \omega^2 - k_z^2$ and
\begin{align}\label{k_rho_integration}
F(q) &= \dfrac{H^{(1)}_n(qR')J_n(qR)}{[(qR)^2 - (p_\rho R)^2]^2},\\
\tilde{F}(q) &= \dfrac{\lim\limits_{q \to p_\rho}[\{(qR)^2 - (p_\rho R)^2\}\times F(q)]}{(qR)^2 - (p_\rho R)^2}.
\end{align}

Contour integration over $k_z$:
\begin{align}\label{k_z_integration}
    &\int_{-\infty}^{+\infty}dk_z \, {\rm K}_z(k_z) \times \dfrac{e^{ik_zl_z}}{k^2_z - (q + i\varepsilon)^2} \nonumber \\ & = \pi i \times {\rm K}_z(q) \times \dfrac{e^{iql_z}}{q},
\end{align}
where $q^2 = \omega^2 - k^2_\rho$ and
\begin{equation}
    {\rm K}_z(k_z) = \left[ \dfrac{\sin(\frac{k_zL}{2})}{\frac{k_zL}{2}}\right]^2
\end{equation}
for TM$_{010}$ and 
\begin{equation}
    {\rm K}_z(k_z) = \left[ \dfrac{ \pi \cos(\frac{k_zL}{2})}{\left(\frac{k_zL}{2}\right)^2 - \left(\frac{p_zL}{2}\right)^2} \right]^2
\end{equation}
for TE$_{011}$.

The exact analytical asymptotic in the adjacent case can be obtained by the approximations under conditions $\varkappa_{A'}L \gg 1$:
\begin{equation}
    \left[ \dfrac{\sin\left(\dfrac{\varkappa_{A'}Lx}{2}\right)}{\dfrac{\varkappa_{A'}Lx}{2}}\right]^2 \simeq \dfrac{2\pi}{\varkappa_{A'} L} \delta (x),
\end{equation}
\begin{equation}
    \left[ \dfrac{ \pi  \cos\left(\dfrac{\varkappa L x}{2}\right)}{\left(\dfrac{\varkappa L x}{2}\right)^2 - \left(\dfrac{\pi}{2}\right)^2} \right]^2 \simeq \dfrac{4\pi}{\varkappa_{A'} L} \delta (x).
\end{equation}

For the case of encapsulated geometry, after integration over $k_\rho$ variable the remaining one dimensional integral over $k_z$ for masses $m_{A'} > \omega$ reads
\begin{align}
    &\int_{-\infty}^{+\infty}dk_z \, {\rm K}_z(k_z) \times e^{ik_zl_z} \nonumber \\
    & \times \dfrac{\pi i}{2} \left[ F(q) - \tilde F(q) \right]_{q = i\sqrt{k_z^2 + \varkappa^2_{A'}}}.
\end{align}

In the case of the TE$_{011}$-mode, the integral with integrand contains $\tilde{F}(i\sqrt{k_z^2 + \varkappa_{A'}^2}) \approx k_z^{-2}$ dominates over the another one with the function $F(i\sqrt{k_z^2 + \varkappa_{A'}^2}) \approx k_z^{-5}$ in the large mass region $m_{A'} \gg \omega$. Therefore, we can neglect the first term. The second one can be analytically calculated in a simple way by the contour integration and Jordan's lemma.

In the case of the TM$_{010}$-mode, the integral with integrand contains $\tilde{F}(i\sqrt{k_z^2 + \varkappa_{A'}^2})$ vanishes due to the multiplayer $(k_z^2 + m_{A'}^2)$ in the function $K_z(k_z)$. Thus, the calculation of asymptotic dependence requires consideration of an integral with $F(i\sqrt{k_z^2 + \varkappa^2_{A'}})$ function. It can be shown, that after using approximation expression,
\begin{equation}
    H^{(1)}_n(qR)J_n(qR) \simeq \dfrac{1}{\pi i R q}, \quad |qR| \gg 1,
\end{equation}
the corresponding integral can be reduced to the integral representation of Macdonald functions $K_\alpha(z)$.

  \section{Asymptotic for large masses
\label{SecAsympt}} 
\begin{enumerate}
    \item Adjacent geometry:
    \begin{enumerate}
        \item TM$_{010}$-mode
        \begin{align}
            \epsilon &= \left[\frac{2 T \, {\rm SNR}}{Q_{\rm rec} (E_0^{\rm em})^2 t}\right]^{1/4} \cdot \left(\dfrac{\pi R L }{4x_{01}}\right)^{1/4} \cdot m_{A'}^1   \\
            & \cdot \left\{\begin{array}{ll}
         1, & \omega d \leq m_{A'}d \ll 1;  \\ 
         \left(\dfrac{8}{\pi m_{A'}d}\right)^{1/4} \cdot  \exp\left(\dfrac{m_{A'}d}{2}\right), & m_{A'}d \gg 1.
         \end{array}\right. \nonumber
        \end{align}
        \item TE$_{011}$-mode (coincides with expression in Ref.~\cite{Berlin:2023mti})
        \begin{align}
        \epsilon &= \left[\frac{2 T \, {\rm SNR}}{Q_{\rm rec} (E_0^{\rm em})^2 t}\right]^{1/4} \cdot\left(\dfrac{\omega^3 L^5}{\pi^5 R^2}\right)^{1/4} \\ 
        & \cdot m_{A'}^{1/2} \cdot\exp\left(\dfrac{m_{A'}d}{2}\right). \nonumber
        \end{align}
    \end{enumerate}
    \item Nested geometry:
    \begin{enumerate}
        \item TM$_{010}$-mode
        \begin{align}
        \epsilon &= \left[\frac{2 T \, {\rm SNR}}{Q_{\rm rec} (E_0^{\rm em})^2 t}\right]^{1/4} \cdot \left[\dfrac{\left(\frac{x_{02}J'_0(x_{02})}{x_{01}J'_0(x_{01})}\right)^2 - 1}{\pi x_{01}}\right]^{1/4}  \\ 
        &  \cdot \left(\dfrac{R_1}{L} \right)^{1/4}\cdot m_{A'}^{1/2} \cdot \exp\left(\dfrac{m_{A'}d}{2}\right).
        \nonumber
        \end{align}
        \item TE$_{011}$-mode
        \begin{align}
        \epsilon &= \left[\frac{2 T \, {\rm SNR}}{Q_{\rm rec} (E_0^{\rm em})^2 t}\right]^{1/4} \cdot\left[\dfrac{\left(\frac{x_{12}J'_1(x_{12})}{x_{11}J'_0(x_{11})}\right)^2 - 1}{\pi x^4_{11}}\right]^{1/4} \cdot\\ 
        &  \cdot\left(\dfrac{\omega^3 R^4_1}{L} \right)^{1/4}\cdot m_{A'}^{1/2} \cdot\exp\left(\dfrac{m_{A'}d}{2}\right).
   \nonumber
        \end{align}
    \end{enumerate}
\end{enumerate}

\color{black}
\bibliography{bibl}

\end{document}